\documentclass[aps,pre,reprint,superscriptaddress]{revtex4-1}
\usepackage{graphicx}
\usepackage{amssymb}
\usepackage{amsmath}
\usepackage{hyperref}
\usepackage{float}
\usepackage{mathtools}

\begin{document}

\title{Dynamical Theory and Cellular Automata Simulations of Pandemic Spread: Understanding Different Temporal Patterns of Infections}

\author{Saumyak Mukherjee}
\affiliation{Solid State and Structural Chemisry Unit, Indian Institute of Science, Bengaluru-560012, India}
\author{Sayantan Mondal}
\affiliation{Solid State and Structural Chemisry Unit, Indian Institute of Science, Bengaluru-560012, India}
\author{Biman Bagchi}
\affiliation{Solid State and Structural Chemisry Unit, Indian Institute of Science, Bengaluru-560012, India}

\begin{abstract}
Here we propose and implement a generalized mathematical model to find the time evolution of population in infectious diseases and apply the model to study the recent COVID-19 pandemic. Our model at the core is a non-local generalization of the widely used Kermack-McKendrick (KM) model where the susceptible (S) population evolves into two other categories, namely infectives (I) and removed (R) where R consists of cured (C) and dead (D) population. This is the well-known SIR model in which we further divide both S and I into high and low risk categories. We first formulate a set of non-local dynamical equations for the time evolution of distinct population distributions under this categorization in an attempt to describe the general scenario of infectious disease progression. We then solve the non-linear coupled differential equations- (i) numerically by the method of propagation, and (ii) a more flexible and versatile cellular automata (CA) simulation which provides a coarse-grained description of the generalized non-local model. In order to account for multiple factors such as role of spreaders before containment, we introduce a time dependent rate which appears to be essential to explain the sudden spikes before the plateau observed in many cases (for example like China). We demonstrate how this generalized approach allows us to handle the effects of (i) time-dependence of the rate-constants of spread, (ii) different population density, (iii) the age ratio, (iv) quarantine, (v) lockdown, and (vi) social distancing. Our study allows us to make certain predictions regarding the nature of spread with respect to several external parameters, treated as control variables. The inherent demographic heterogeneities in a population distribution tend to become magnified in the disease prediction due to the non-linearity in the system of equations. Analysis of the model clearly shows that due to the strong heterogeneity in the epidemic process originating from the distribution of initial infectives, the theory must be local in character but at the same time connect to a global perspective.
\end{abstract}

\maketitle

\section{Introducntion}

The extremely unfortunate appearance of the coronavirus disease (COVID-19 or SARS-Cov2) has taken the entire world aback.\cite{Sohrabi2020} On March 11th, 2020 the world health organisation (WHO) has declared COVID-19 a pandemic. During the last one and half months, the situation has progressively become worse, making the world forcing a challenge we did not face in recent times. At this juncture, it is not clear how we shall come out of this misfortune and its aftereffects. Under this extremely terrifying and frustrating situation we need to look for solutions that might help in effective planning and policymaking. This is extremely demanding especially for countries with limited resources.

Healthcare professionals, sociologists and politicians strive to find out an optimised plan to cure and contain the disease. On the other hand, scientists attempt to understand and analyse different aspects of this problem with respect to different natural and imposed conditions to predict the future of the pandemic.\cite{Petropoulos2020,Yamana2019} A molecular level microscopic understanding of the virus has already emerged that has accelerated the process of vaccine preparation.\cite{Wrapp2020,Chen2020} However, detailed theoretical and simulation based studies from the demographic and topographic viewpoints are still being pursued.

There are several mathematical models which have been employed in the context of epidemic modelling.\cite{Siettos2013} For example, Kermack-McKendrick (KM) model has been used extensively to study the spread of infectious diseases like measles, small pox etc.\cite{Kermack1927,Daley2001,Diekmann1995} At the core of this model lies a system of three coupled differential equations for susceptible (S), infected (I) and removed (R) (cured and dead) populations, that is, the famous SIR model [Eq. \ref{eq1}]. At the onset of an epidemic S becomes I and I eventually becomes R, but R can never become S or I because of acquired immunity. 
 
\begin{equation}
 \begin{split}
  \frac{dS}{dt}& =-k_{S\rightarrow I}SI\\
  \frac{dI}{dt}& =k_{S\rightarrow I}SI-k_{I\rightarrow R}I\\
  \frac{dR}{dt}& =k_{I\rightarrow R}I
 \end{split}
 \label{eq1}
\end{equation}

Eq. \ref{eq1} describes the three coupled non-linear differential equations of the KM model where $k_{S\rightarrow I}$ is the rate of infection and $k_{I\rightarrow R}$ is the rate of removal (recovery and death). In the conventional SIR model $k_{S\rightarrow I}$ and $k_{I\rightarrow R}$ are written as $\beta$ and $\gamma$ respectively. In principle the rate constants should be time and space dependent, that is, non-local in nature. However, this model is applicable for a homogeneous population distribution and mass transmission at a large scale.\cite{Daley2001}

\begin{figure}[H]
 \centering
 \includegraphics[width=3.2in,keepaspectratio=true]{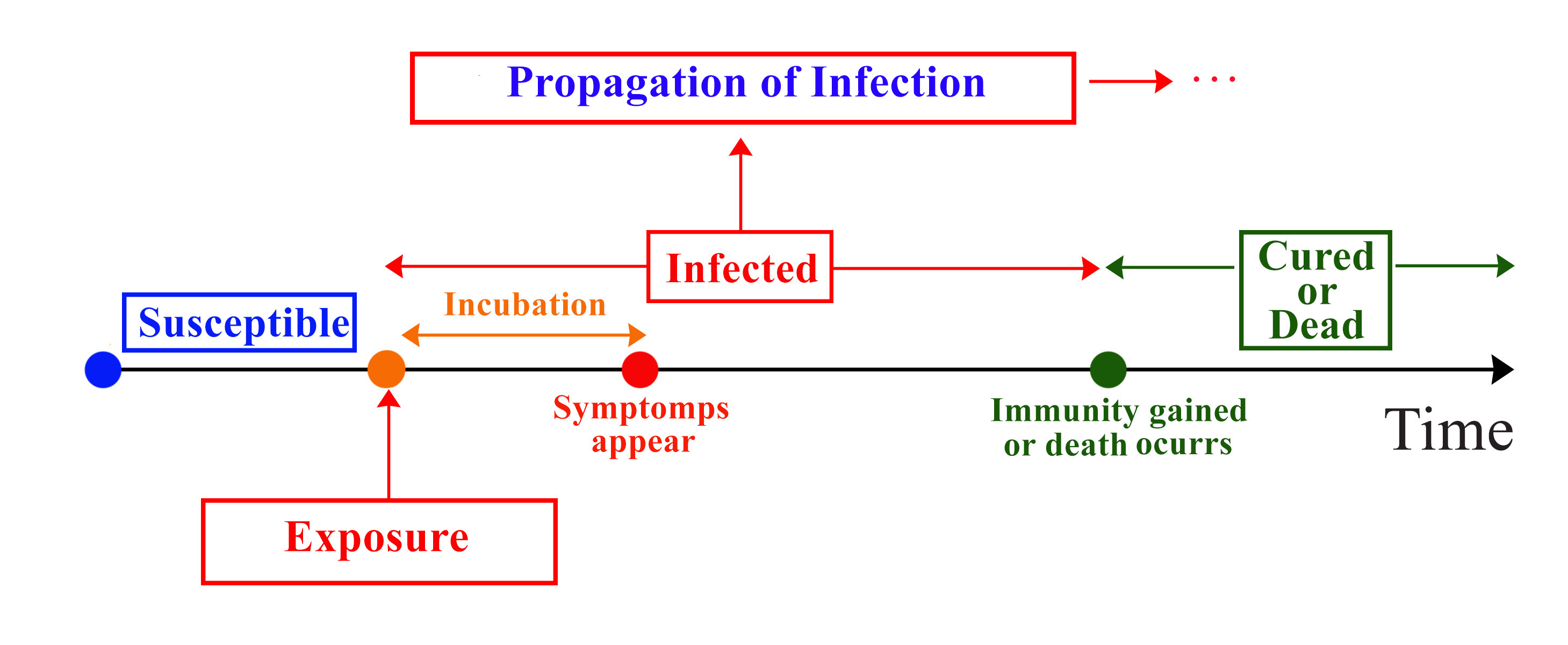}
 \caption{Schematic representation of the propagation of an infectious disease. A susceptible person (S) can become infected (I) by exposure to the virus. The virus incubates for a certain time inside the body and the symptoms appear after the incubation period. However for certain individuals symptoms might not appear (known as asymptomatic carriers). After the infection period a patient might recover (and gain immunity) or die.}
 \label{fig1}
\end{figure}

In Fig. \ref{fig1} we provide a schematic timeline of the propagation of the infection for an individual. We note that this situation is exactly similar to that of a consecutive chemical reaction like radioactive decay kinetics.\cite{steinfeld1989chemical}

There are several existing modelling and simulation studies which have showed a way to the policy makers and strategists to take proactive measures.\cite{Kucharski2020,Ming2020,Mandal2020,Iwata2020} Recently a number studies have emerged which have both qualitatively and semi-quantitatively demonstrated the advantages of social distancing and lockdown.\cite{Organization2020,Prem2020} Few of them have attempted to predict the optimum duration of the lockdown and social distancing policies.\cite{Singh2020} However, such mathematical model based studies fail in many aspects because of the complexity of the real life situation, stochastic nature crowd dynamics, herd mentality of humans, and inherent time-dependence of rate processes. Hence the predictive power of these models becomes limited.

In this work we employ a modified and generalised version of the KM model and incorporate several other parameters to mimic some real life social requirements with certain degree of specificity. Countries like Italy, where the average age is higher and population density is low, are expected to respond differently compared to countries like India where the average age is far lower and population density is high. Moreover the different weather conditions and of tropical countries might play a role.\cite{Casanova2010} However, the sensitivity of COVID to the weather and immunity against other diseases is under investigation and yet to be confirmed. All these situations should, in principle, modulate the transmission probabilities and rate of spread.

We organise the rest of the paper as follows. In section II we detail the mathematical modelling and cellular automata simulation protocols. Section III contains the numerical results and analyses. This section is further divided into several subsections that show the effect of different control parameters (for example, probability of transmission, ratio of young and old people, population density etc.) on the total fraction of S, I and R from CA simulations. We note that the present study and related mathematical model based studies might exhibit high sensitivity to the parameter space and underlying assumptions. Such studies are aimed to help understand the strategic moves and do not contain clinically implementable results.

\section{Mathematical Modelling: Time-Dependent Rate Constant}

Our model is a generalization of the celebrated Kermack-McKendrick (KM) model or SIR theory for epidemic modelling.\cite{Kermack1927, Anderson1979, Jones2009, Skvortsov2007} Accordingly, we divide the entire population of a region into the following categories.
1.	Susceptible (S): These are the initially healthy fraction of the population who are susceptible to get infected. 

2.	Infectives (I): These are 

3.	Removed (R): These people either get cured from the disease or die. The cured population develop the required anti-body in their system so that they are safe from a second chance of infection. We neglect the re-infection scenario which is extremely rare.

Based on this classification, the time total population ($N_{tot}$) can be written as a summation of the numbers people in each category (Eq. \ref{eq2}).
\begin{equation}
 N_{tot} = N_S(t) + N_I(t) + N_R(t)
 \label{eq2}
\end{equation}

\begin{figure}[h]
 \centering
 \includegraphics[width=2.8in,keepaspectratio=true]{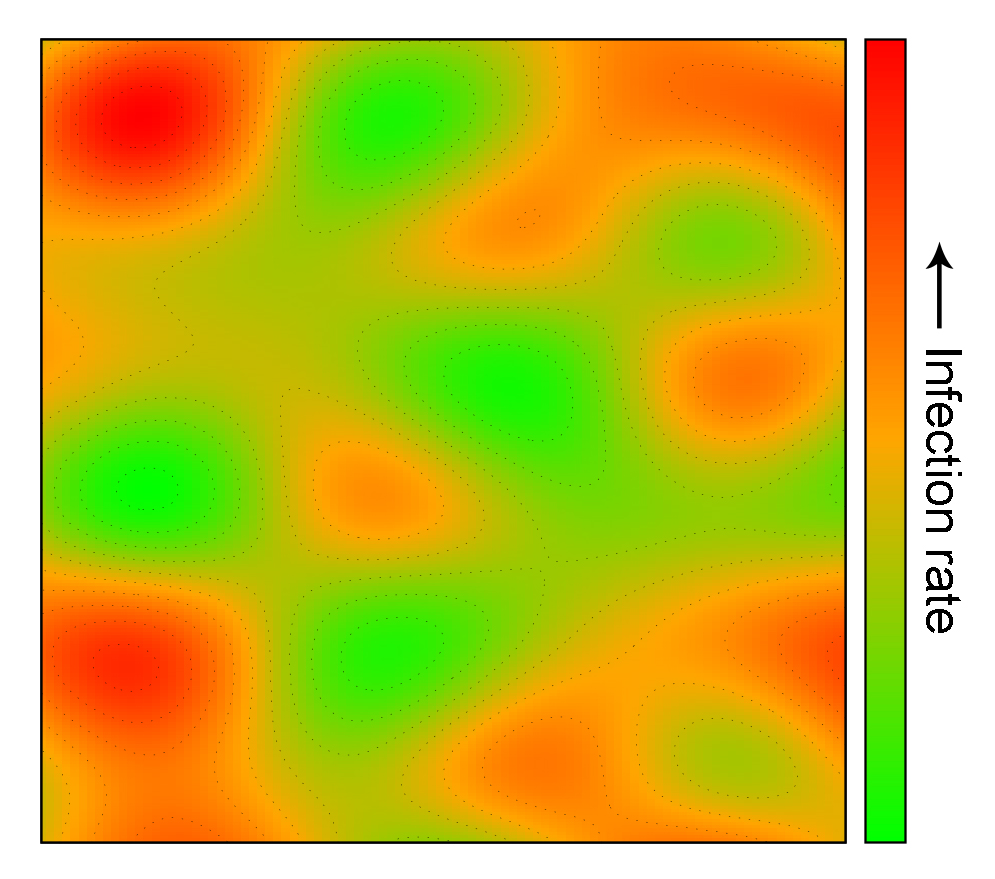}
 \caption{A schematic contour map to demonstrate the non-local behaviour and time-dependence of the rate of disease spread. Here, red denotes regions with high rate constants while green represents lower rates. The pattern portrayed in this map is time-dependent and may change accordingly.}
 \label{fig2}
\end{figure}

Here, $N_X(t)$ is the fraction of population in each category (X = S, I, R). For any pandemic, the rate of conversion from S to I is nonlinear. There is an inherent spatio-temporal dependence of the rate constants. We generalize the conventional SIR model by introducing this dependence, by expressing the rate constants as functions of time and space. This is schematically demonstrated by a contour map in Fig. \ref{fig2}. Here, red denotes regions with higher rate of disease spread, while green denotes a lower rate. The patterns in the map experience temporal evolutions denoting the time-dependence of the rate constants. Hence, a realistic model of the spread of an infectious disease requires rate constants that depend both on time and space. 

We first formulate the non-local space-time dependent generalisation of the KM equations [Eq. \ref{eq3} - \ref{eq6}] where the rates (k) as are expressed as integrations over space and time.

\begin{equation}
 \frac{\partial}{\partial t}N_S(\mathbf{r},t) = -N_S(\mathbf{r},t)\int d\mathbf{r}^\prime dt^\prime k_{S\rightarrow I}(\mathbf{r}-\mathbf{r}^\prime)N_I(\mathbf{r}^\prime,t^\prime)
 \label{eq3}
\end{equation}

\begin{equation}
\begin{split}
 \frac{\partial}{\partial t}N_I(\mathbf{r},t) = &N_S(\mathbf{r},t)\int d\mathbf{r}^\prime dt^\prime k_{S\rightarrow I}(\mathbf{r}-\mathbf{r}^\prime)N_I(\mathbf{r}^\prime,t^\prime) \\
&- (k_{I\rightarrow C} + k_{I\rightarrow D})N_I(\mathbf{r},t)
 \label{eq4}
 \end{split}
\end{equation}

\begin{equation}
 \frac{\partial}{\partial t}N_R(\mathbf{r},t) = k_{I\rightarrow R} N_I(\mathbf{r},t)
 \label{eq5}
\end{equation}

Further integration of $N_X(t)$ over all space gives us a set of 3 non-local equations which represent a more general model with time-dependent mean field description of rate constants. The solution of these equations is however highly non-trivial. This can be simplified if we neglect the non-local nature of the disease spread. Hence we can reduce the above equations to the following forms (Eq. \ref{eq6}-\ref{eq8}).

\begin{equation}
 \frac{dN_S(t)}{dt} = -k_{S\rightarrow I}N_S(t)N_I(t)
 \label{eq6}
\end{equation}

\begin{equation}
 \frac{dN_I(t)}{dt} = k_{S\rightarrow I}N_S(t)N_I(t)-k_{I\rightarrow R}N_I(t)
 \label{eq7}
\end{equation}

\begin{equation}
 \frac{dN_R(t)}{dt} = k_{I\rightarrow R}N_I(t)
 \label{eq8}
\end{equation}

with the constraint

\begin{equation}
 \frac{dN_{tot}}{dt} = \frac{d}{dt}[N_S(t)+N_I(t)+N_R(t)] = 0
 \label{eq9}
\end{equation}

Here, $k_i$ is the rate constant for the $i^{th}$ process. The arrow ($\rightarrow$) in the subscript of the rate constant denotes the direction of the respective process. One important parameter is the \textit{reproductive ratio} ($R_0$) as defined in Eq.\ref{eq10}.\cite{Holmes1997}

\begin{equation}
 R_0=\left(\frac{k_{S\rightarrow I}}{k_{I\rightarrow R}}\right) N_S
 \label{eq10}
\end{equation}

The value of $R_0$ determines the number of secondary infections from a primary infection.\cite{Hartfield2013} For COVID-19 the value average of $R_0$ has been determined as approximately 2.2.\cite{Zhang2020} We consider the time dependence of $R_0$ in our model. Accordingly $k_{S\rightarrow I}$ evolves over time but $k_{I\rightarrow R}$ remains same as it depends on the nature of the disease.

However, the position of the individuals in the system is also important. For example, the factors like social distancing, lock down, quarantine etc. are explicitly position dependent. KM model is strictly applicable for homogeneous population distribution and lacks this space-dependence.\cite{Daley2001}

In this work we adopt two different approaches for reasons discussed below. The first is a quasi-analytical method where we perform numerical integration to solve the local coupled equations. This takes into account the time dependence of rate constants. This does not involve simulations, but also ignores non-locality. In the second approach we carry out cellular automata simulations, where the non-locality is partly implemented as necessary to address features like lockdown and quarantine. 

\section{Stochastic Cellular Automata Simulation}

The Eqs. \ref{eq6}-\ref{eq8} are approximations of the more general non-local partial differential equations \ref{eq3}-\ref{eq5}. In a real-world scenario, the non-local description may often become important in determining the fate of a pandemic in a given geographical region. In such a case, the population parameters are space-dependent. Moreover, the rate constants also have a spatial distribution. Hence, solutions of these equations are highly non-trivial. As a means of incorporating this non-local behaviour of the system into our model, we seek the help of stochastic cellular automata (CA) simulations. CA simulation techniques are often used to model several physical phenomena.\cite{Hollingsworth2004, Seybold1998, Wolfram1983, Bartolozzi2004, Soares-Filho2002, Goltsev2010, Almeida2011} Unlike the mathematical model, CA simulations present a more microscopic view into the problem and directly establish a physical map of the disease-spread. Moreover, the generality of this scheme lies in the fact that we incorporate several parameters, individual exploration of which allow us to specifically understand the role of different factors that are important for the spread of a disease. These parameters can be classified into three primary categories:

1. Region specific: These are the parameters that define the geographic and demographic characteristics of the region in which the disease is spreading. These include (a) population density, (b) fraction of initial healthy and infected individuals, and (c) age distribution in the population.

2. Disease specific: These parameters are characteristics of the disease that has caused the epidemic (viz. COVID-19, SARS, Ebola, Cholera, Influenza etc.). This includes (a) transmission rate, (b) mortality rate, (c) probability of recovery/death according to age, and (d) incubation period of the virus.

3. Disease control: The parameters that are important for the control of the spread of the disease are the fractions of the population abiding by the norms of (a) social distancing and (b) quarantine/lock down as put forward by the government.

We use the Moore definition \cite{Kier2005} to define the neighbourhood of an individual. It should be noted that the von Neumann neighbourhood \cite{Kier2005} definition neglects the isotropic nature of infection in a disease transmission. Several previous CA simulations used to model epidemic spread have also used Moore neighbourhood.\cite{Fu2003, White2007, Sirakoulis2000} For a pandemic, it is often difficult to obtain the rates of the processes (S to I or I to C/D) with respect to the individual factors mentioned above. This may be due to lack of proper information or an inherent difficulty in addressing a particular factor. However, from the average available data, it is easier to estimate probabilities pertaining to the concerned factors. Hence, rather than modelling the CA simulation based on rate constants, we employ a probabilistic approach. The salient features of our simulation are noted below.

1. We divide the “removed” (R) category of the SIR model into cured (C) and dead (D). Hence an SICD model of stratified population is used in these simulations. Further, the susceptibles are classified into high risk and low risk categories, depending on their probability of getting infected. 
	
2. We start with a stretch of land with fractions of it covered by healthy/susceptible (S) people and infected (I) people. The total area for the simulation is given by an $N_1\times N_2$ matrix of boxes. Each cell (point) is either a vacant land (V) or a human being. The positions of the individuals are initialized randomly within the given area. The initial fractions of S and I determine the population density of the region. Note that an initial non-zero fraction of I is important for this simulation as it acts as the trigger for the spread of the disease. Moreover, there are nor cured (C) or dead (D) people in the beginning. 
	
3. When an infective comes in the neighbourhood of a susceptible, the latter gets infected with a certain transmission probability. This probability is time dependent and is defined by a given function (logistic/exponential/oscillatory). For individuals with low susceptibility to infection the value of transmission probability is lower. This fraction of people include individuals who practise social distancing, have high disease resistance or are extremely cautious about their surroundings.
	
4. Since infectives are initially present in the population, a fraction of the susceptible group is assigned to be quarantined (or locked down) from the beginning. This is achieved by fixing them at their respective positions throughout the simulation. It is assumed that these people cannot be infected as they cannot come in contact with the infectives.
	
5. An age between 0 and 100 is assigned to each individual in the entire population at the beginning of the simulation. A threshold age is taken to determine whether an individual is young or old. Accordingly, a fraction of young people is defined. In this simulation the young individuals represent the people with higher immunity. This has been seen in several cases.\cite{Remuzzi2020,Ruan2020,Wu2020}
	
6. In a given step propagation in the simulation, an infective can either die or recover or remain infected. A probability determines whether the individual continues to remain infective of not. If not, we define two recovery probabilities for the young and the old. The latter is much lower than the former. These 3 probabilities together determine the duration in which an individual remains infected and is potentially dangerous for the susceptible fraction of the population. This time duration relates to the incubation period of the virus.

7. A recovered/cured person becomes immune to further infection.

In Table \ref{tab1} we provide the definitions of the notations related to the above description used in our CA simulations.

\begin{table}[ht]
\caption{Definitions of the notations in this work.}
 \centering
 \begin{tabular}{|c|l|}
  \hline
  Notation & Definition\\
\hline
\hline
  $F_X^0$ & Initial fraction of individuals\\& in the category X (S, I, C, D)\\& with respect to the available land area\\
  \hline
  $F_X$ & Time dependent fraction of individuals\\& in the category X (S, I, C, D)\\& with respect to the total initial population\\
  \hline
  $F_Y$ & Fraction of the S strata consisting of\\& young individuals (people with higher\\& immunity in general) The term young is\\& defined by a threshold age A.\\
  \hline
  $P_{Tr}$ & Probability of transmission of\\& the disease when an infective comes\\& in contact with a susceptible individual\\
  \hline
  $P_R^Y$ & Probability of recovery\\& of a young infective \\
  \hline
  $P_R^O$ & Probability of recovery\\& of an old infective \\
  \hline
  $P_I$ & Probability that an infective remains\\& infected without recovery or death\\& in a simulation time step\\
  \hline
  $F_Q$ & Fraction of susceptibles\\& in quarantine or lock-down\\
  \hline
  $F_SD$ & Fraction of people with\\& low susceptibility to infection\\& (SD denotes social distancing)\\
  \hline
  $P_{Tr}^{SD}$ & Transmission probability of disease\\& for an individual within\\& the fraction $F_{SD}$\\
  \hline
  $F^{Saturation}$ & The value at which a given\\& time-dependent fraction gets saturated\\
  \hline
  $F_I^{max}$ & The height (fraction) of the maxima of $F_I$\\
  \hline
  $t_I^{max}$ & The position (time) of the maxima of $F_I$\\
  \hline
  \end{tabular}
 \label{tab1}
\end{table}

We employ all these conditions and run our simulations for a given number of steps (N). It should be noted that the time unit ($\Delta t$) is at our liberty to choose. Hence, comparison with real data is important for prediction of the time-scale of the epidemic. Additionally, this simulation can be applied to a wide range of diseases and geographical regions by using the right parameters. 

\begin{figure}[h]
 \centering
 \includegraphics[width=3in,keepaspectratio=true]{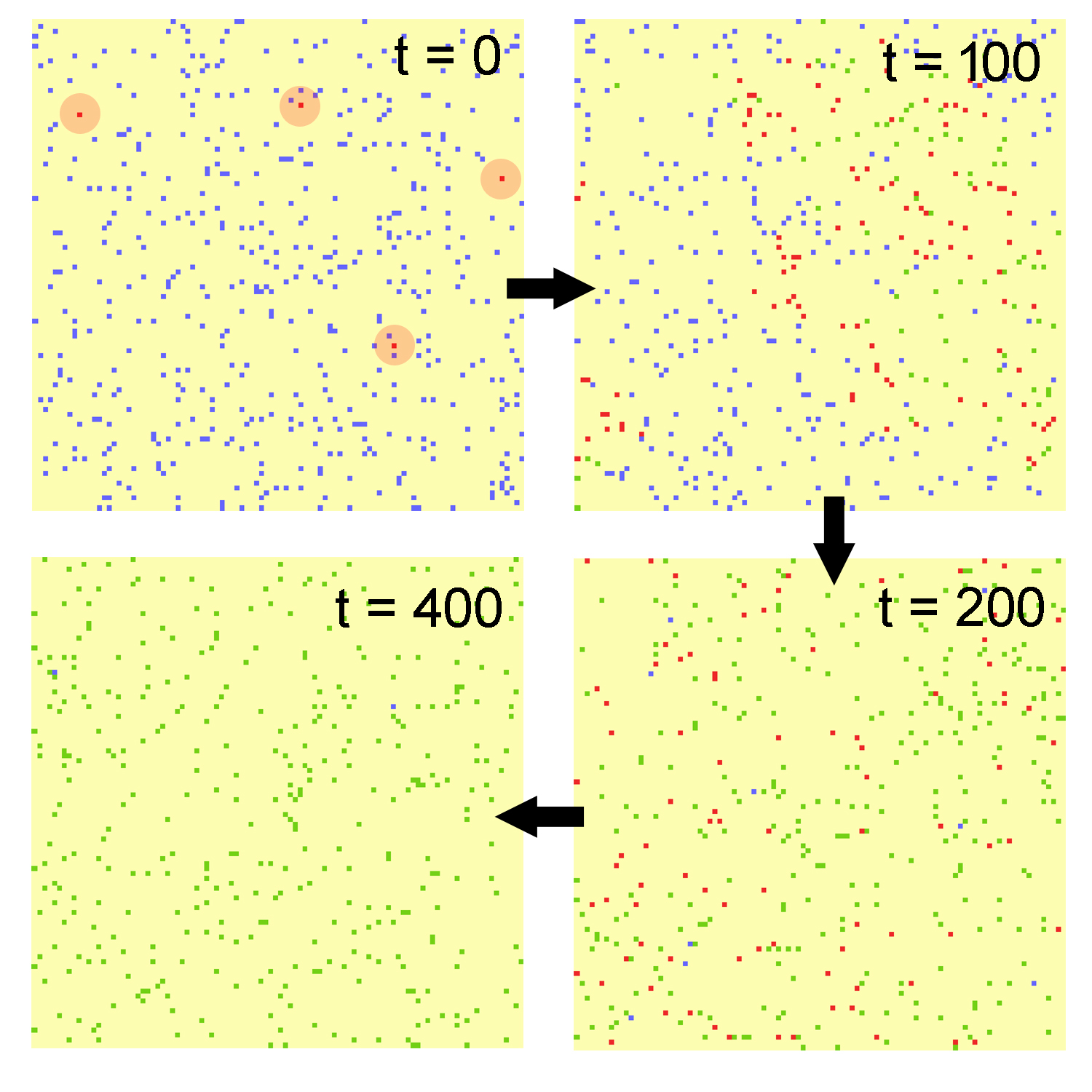}
 \caption{Snapshots of the population map at different time steps from CA simulation according to the protocol explained above. The colours blue, red and green denote susceptible, infective and cured respectively. The dead are taken out of the population. Starting with a very few infectives (marked at $t=0$) within a society dominated by susceptibles, the disease spreads throughout the population. In the final frame most of the population is infected and is either cured or dead.}
 \label{fig3}
\end{figure}

The population mapped out by a representative cellular automata simulation is shown in Fig \ref{fig3}. The initial population consists of 5 \% of the total available land (light yellow) covered with susceptibles (blue) and 0.1 \% by infectives (red). There are only four infectives at $t=0$ (marked by transparent red circles). With the advance of time, the infectives transmit the disease to the susceptibles. The infected people can either recover (green) of die (removed from the population). At the end, most of the population is infected and they have either recovered or died.
 
\section{Results and Discussion}

From the theoretical layout discussed above, we can get the time evolution of the four categories in a population (susceptibles, infectives, cured and dead) for given rates of the respective processes. Hence, from the available data, if we are able to extract the said rates of conversion, we can predict the future scenario of the COVID-19 pandemic for different regions in the world. Let us take the example of India. The case in India was reported on $30^{th}$ January, 2020. From then, the count of infection has seen an exponential increase reaching 24,434 on $24^{th}$ April, 2020. The death toll has reached 780 the first one being reported on $12^{th}$ March (source: www.covid19india.org). However, the number of patients recovering also increases exponentially, although at a lower rate as compared to the infection. From the data we extract the rate constants as given in Table \ref{tab2}.

\begin{table}[H]
\caption{Values of rate constants of respective processes in India (up to $24^{th}$ April 2020).}
 \centering
 \begin{tabular}{|c|c|}
  \hline
  Rate constant & Value ($day^{-1}$)\\
\hline
\hline
  $k_{S\rightarrow I}$ & 0.116\\
  \hline
  $k_{I\rightarrow C}$ & 0.016\\
  \hline
  $k_{I\rightarrow D}$ & 0.004\\
  \hline
 \end{tabular}
 \label{tab2}
\end{table}

We use these values to estimate $R_0$ to be $\sim$6. Nevertheless, the value of $R_0$ evolves over time as the people and government take protective measures. As mentioned above the span of the values of rate constants can depend on region, average age of population and other important factors. 

\subsection{Solution of SIR model with time-dependent rate}

\begin{figure}[H]
 \centering
 \includegraphics[width=3.4in,keepaspectratio=true]{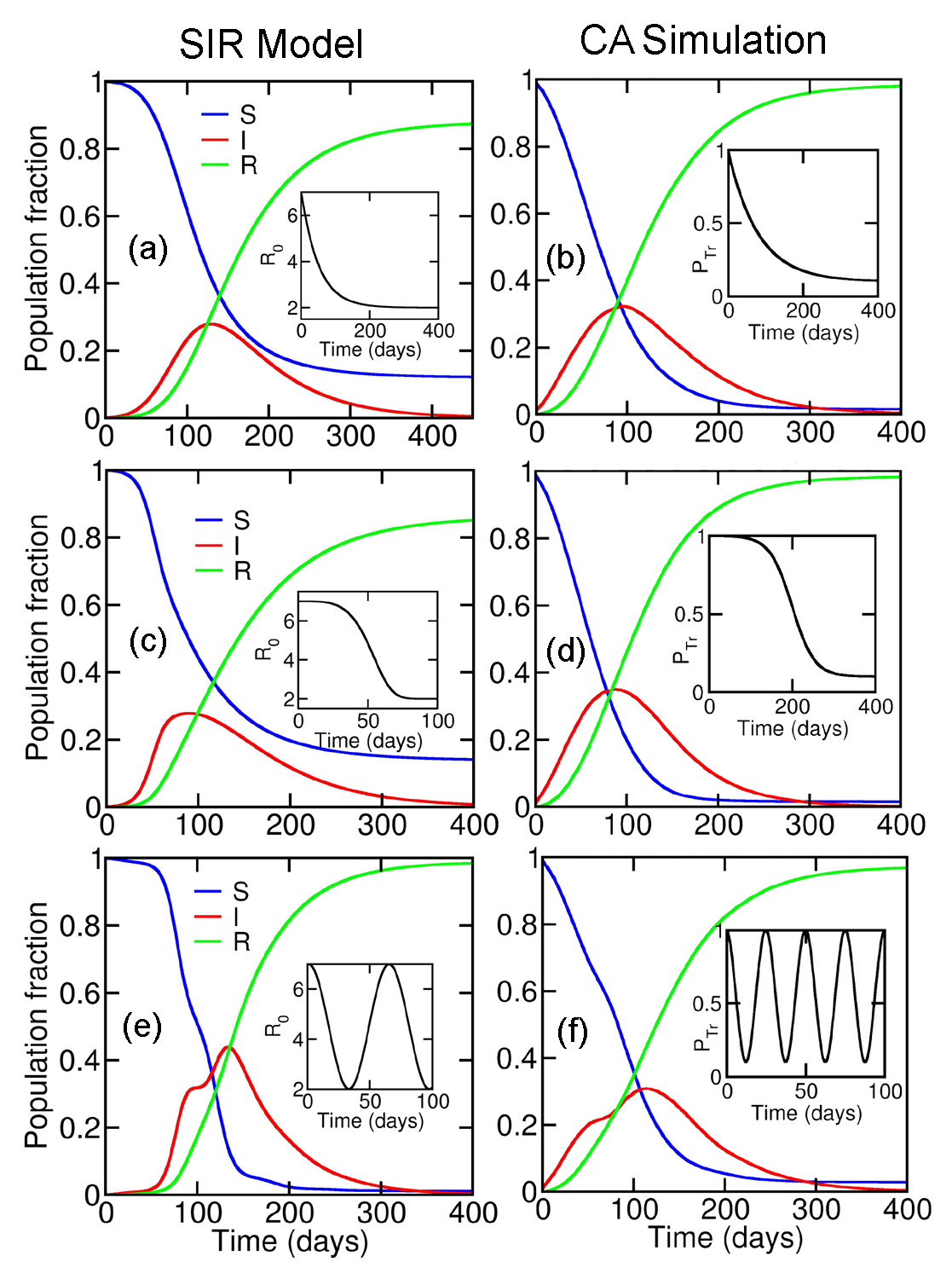}
 \caption{Time evolution of the S (susceptible), I (infective) and R (removed) fractions of the population as observed from the generalized SIR mathematical model (left column) and stochastic cellular automata simulation (right column). These time dependencies are obtained by using different functional forms to define the time dependent rate ($R_0$) in the mathematical model and transmission probability in the simulations ($P_{Tr}$). The functions used are exponential (a, b), logistic (c, d) and square cosine (e, f). The temporal variations of the rates/probabilities are shown in the insets of each graph.}
 \label{fig4}
\end{figure}

We numerically propagate and solve the SIR model [Eq.\ref{eq6}-\ref{eq8}] and obtain the time evolution of $N_S$, $N_I$, and $N_R$ [Fig. \ref{fig4}a, \ref{fig4}b and \ref{fig4}d]. We implement three different kinds of time dependence in $R_0$ so that it varies between 7.0 to 2.0 – (i) exponentially decaying [Eq.\ref{eq11}], (ii) a sigmoidal/logistic function [Eq.\ref{eq12}], and (iii) periodic [Eq. \ref{eq13}]. We choose the upper and lower limits keeping in mind that the infection rate is higher in the beginning and eventually comes down to the average value at longer times. The time evolution plots are supplied in the insets of Fig. \ref{fig4}.

\begin{equation}
 R_0(t)=2.0+5.0exp\left(-\frac{t}{50.0}\right)
 \label{eq11}
\end{equation}
\begin{equation}
 R_0(t)=2.0+\frac{5.0}{1+exp\left[\left(\frac{t}{50.0}\right)^4\right]}
 \label{eq12}
\end{equation}
\begin{equation}
 R_0(t)=2.0+5.0cos^2\left(-\frac{t}{20.0}\right)
 \label{eq13}
\end{equation}

From Fig. \ref{fig4} we can see that the behaviour of the evolution of S, I and R changes with the changing nature of the time-dependence of infection rate. Interestingly, with the periodic variation of $R_0$ we can see a bimodal nature in the I vs time plot. This mimics the data from China. The periodic variation in the rate might happen when after a certain time the preventive measures are relaxed. 

We verify the evolution of the stratified population as obtained from the mathematical model by our stochastic cellular automata simulations. The results are presented in the right hand side column of Fig. \ref{fig4}(b, d, f). Three functional forms of the transmission probability ($P_{Tr}$) are used to define its time dependence. These are exponential [Eq. \ref{eq14}], logistic [Eq. \ref{eq15}] and square cosine [Eq. \ref{eq16}] functions as used in the previous mathematical analysis.

\begin{equation}
 P_{Tr}(t)=0.1+0.9exp\left(-\frac{5t}{T}\right)
 \label{eq14}
\end{equation}
\begin{equation}
 P_{Tr}(t)=0.1+\frac{0.9}{1+exp\left[\left(-\frac{t-(T/2)}{T/15}\right)\right]}
 \label{eq15}
\end{equation}
\begin{equation}
 P_{Tr}(t)=0.1+0.9cos^2\left(-\frac{10t}{T}\right)
 \label{eq16}
\end{equation}

Here, T = total number of simulation steps. The value of $P_{Tr}$ is varied from 1.0 to 0.1 with the time evolution of the simulation according to these functions. The numerical values of the parameters mentioned in the equations are obtained by trial and error to compare with the mathematical solution of the SIR model. Under comparable set of parameters, we find similar patterns of temporal evolution of population from both the generalized SIR model and CA simulation. The striking similarity of the results obtained from the two formalisms helps in validation of our CA simulation protocol. The following sets of parameters (Tab. \ref{tab3}) were used in each of the three simulations.

\begin{table}[h]
\caption{Values of the set of parameters used in the CA simulations mentioned above.}
 \centering
 \begin{tabular}{|c|c|}
  \hline
  Notation & Definition\\
\hline
\hline
  $F_S^0$ & 0.05\\
  \hline
  $F_I^0$ & 0.0005\\
  \hline
  $F_Y$ & 0.6\\
  \hline
  $P_R^Y$ & 0.9\\
  \hline
  $P_R^O$ & 0.6\\
  \hline
  $P_I$ & 0.99\\
  \hline
  $F_Q$ & 0.0\\
  \hline
  $F_SD$ & 0.5\\
  \hline
  $P_{Tr}^{SD}$ & 0.3\\
  \hline
  \end{tabular}
 \label{tab3}
\end{table}

\begin{figure}[H]
 \centering
 \includegraphics[width=3.2in,keepaspectratio=true]{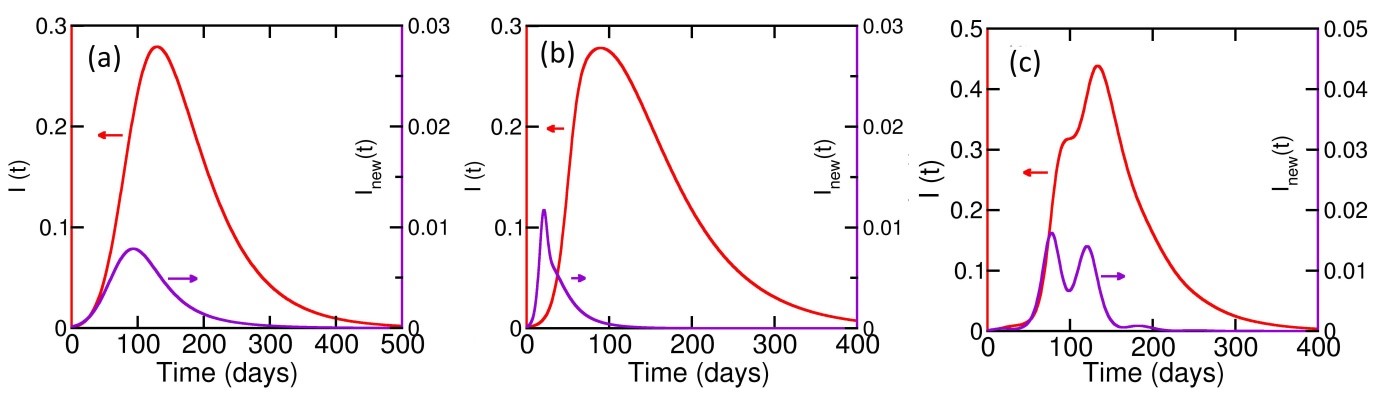}
 \caption{The time evolution of total active infected population [red trace, I(t)] and newly infected population [purple trace, $I_{new}(t)$] for three different kinds of time dependent $R_0$ namely (a) exponentially decaying [Eq. \ref{eq11}], (b) a sigmoidal/logistic function [Eq. \ref{eq12}], and (c) periodic [Eq. \ref{eq13}]. The peak of $I_{new}(t)$ always appears before the peak of I(t). For a periodic time-dependence we see a second peak in $I_{new}(t)$ which is known as a second wave. These results are obtained from the numerical solutions of the SIR model.}
 \label{fig5}
\end{figure}

Another important and relatively less discussed quantity is the fraction of new infection ($I_{new}$). In Fig. \ref{fig5} we show the time evolution of total active infected population [I(t)] and newly infected population [$I_{new}(t)$] for three different kinds of time dependent $R_0$. These analyses show that the maxima in the new cases peak before the maxima of the total number of active infections. We observe a second peak in $I_{new}(t)$ when $R_0$ varied periodically with time. This can be attributed to the \textit{second pandemic wave}.\cite{Barro2020} One is free to incorporate other kinds of temporal variation of the rate constants to explain the date of a specific reigion.

\subsection{Observations from CA simulations: Dependence of epidemic outcome on different parameters}

Here, we study the temporal evolutions of the different categories of the society as functions of the different parameters defined in the CA simulation. All the results presented here are averaged over 100 individual simulations.

\subsubsection{Initial fraction of susceptibles and infectives}

The spread of any infectious disease shows strong sensitivity towards the population density of a region. It also depends on the fraction of susceptible (healthy) and infected population already present in that region. We first vary $F_S^0$ in the range 0.01 to 0.50 by keeping $F_I^0$ fixed at 0.001. The meanings of the notations are given in Table \ref{tab1}.

We keep an order of magnitude difference between $F_S^0$ and $F_I^0$. In all these simulations we have not imposed any social distancing or other proactive measures. In the real scenario the number of infected people in a region would depend on the delay of closure of geographical border to prevent immigration. We note that increase in either of $F_S^0$ or $F_I^0$ would result in an increased population density as the area in the CA simulation has been kept fixed. 

\begin{figure}[h]
 \centering
 \includegraphics[width=3in,keepaspectratio=true]{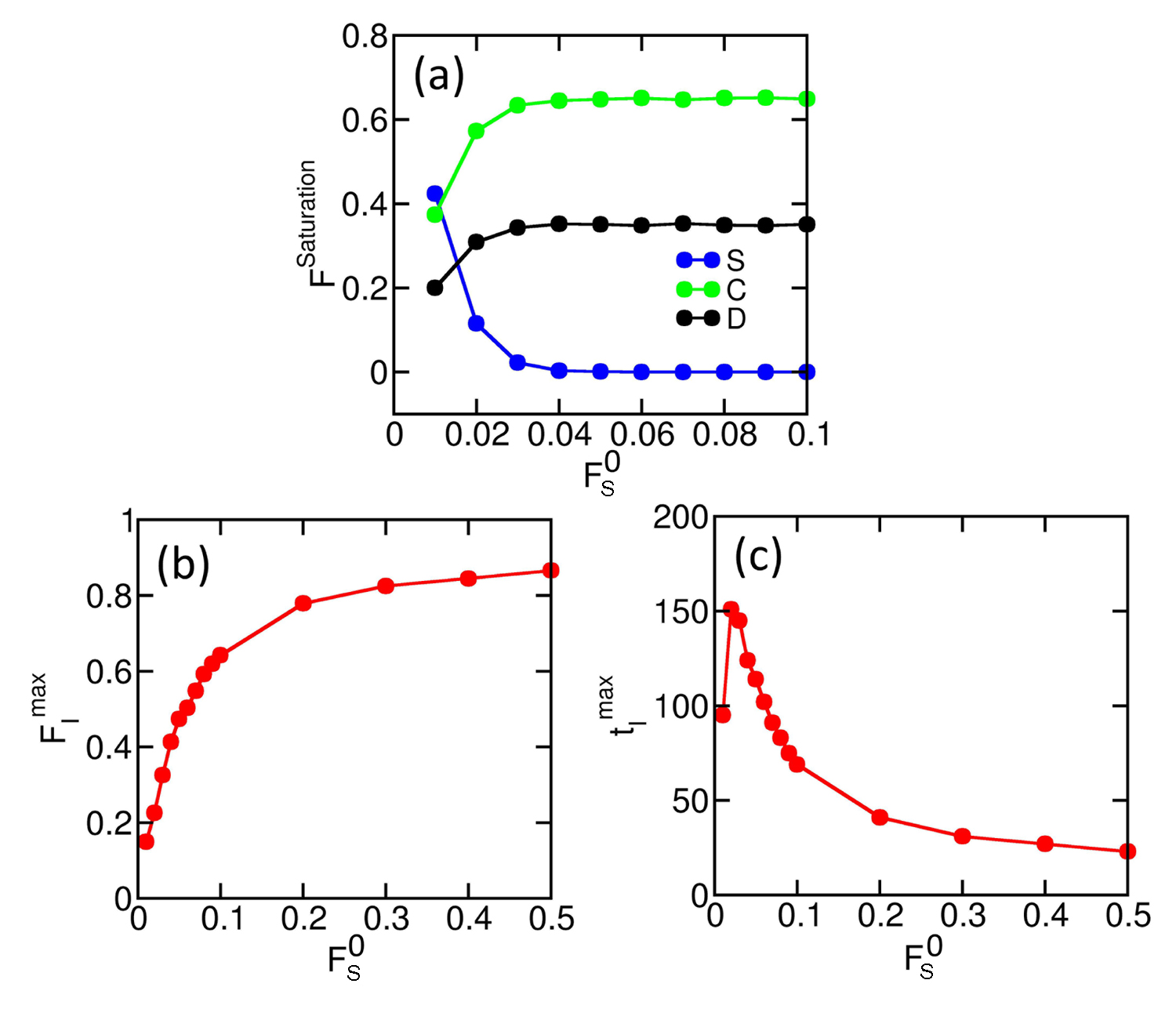}
 \caption{(a) Dependence of the saturated values of the S, C and D population fraction on $F_S^0$. (b) Variation of the peak maxima of $F_I$ vs. time plot with $F_S^0$. (c) Variation of the position of the peak maxima of the same. These temporal patterns are obtained from CA simulations.}
 \label{fig6}
\end{figure}

In Fig. \ref{fig6}a, we plot the saturated (or steady state) values from the plots of $F_S$, $F_C$ and $F_D$ against $F_S^0$. The time evolution is given in the supporting information [Fig. S1]. We see a surprising non-linear behaviour. When $F_S^0$ is higher than 0.04 the whole population becomes infected after a certain time. However, the number of deaths does not vary substantially. \ref{fig6}b and \ref{fig6}c respectively show the variations of the peak height ($F_I^{max}$) and peak position ($t_I^{max}$) against $F_S^0$ from the $F_I$ vs time plot. It is clear that the fraction of infected population increases with increasing $F_S^0$. The time at which the infection becomes a maximum decreases with increasing $F_S^0$. Both the scenarios have negative impact on a society and its healthcare resources.

\begin{figure}[h]
 \centering
 \includegraphics[width=3in,keepaspectratio=true]{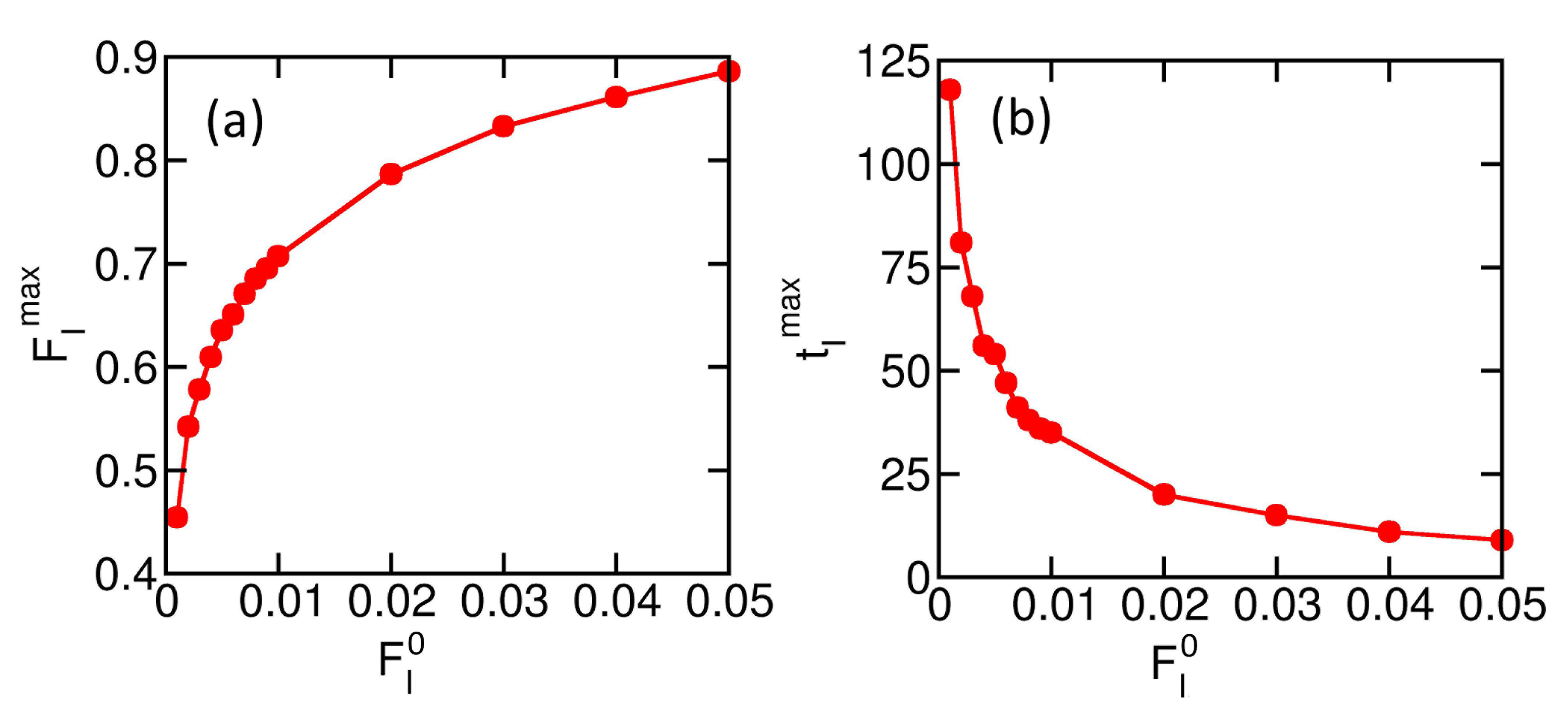}
 \caption{Variation of the (a) height and (b) position of the peak maxima of $F_I$ vs. time graph with $F_I^0$ as obtained from CA simulations.}
 \label{fig7}
\end{figure}

Next we vary $F_I^0$ for a fixed value of $F_S^0=0.05$ and plot the time dependence of S, I, C and D [Fig. S2]. In Fig. \ref{fig7} we perform similar analyses as those in Fig. \ref{fig6}. We see that the infected peak height increases and shifts to lower values in the time axis in a non-linear fashion with increasing $F_I^0$. These results clearly demonstrate the importance of an early ban on international or domestic travels which directly increases the value $F_I^0$.

\subsubsection{Fraction of young population}

The prognosis of patients with infectious diseases such as 
-19, Spanish flu etc. often shows strong sensitivity towards the age of the patient. For example the novel coronavirus has proven to be fatal for older population, especially for people above 50 years and beyond. Importantly, the age distribution is widely different for different countries. For European countries like Italy the average age is way higher than that in India. Keeping this in mind, we vary the fraction of younger population ($F_Y$) from 0.1 to 0.9 in our simulations while the other parameters are fixed. By the term young we indicate the population below 50 years of age especially in the context of COVID-19. 

\begin{figure}[h]
 \centering
 \includegraphics[width=2.5in,keepaspectratio=true]{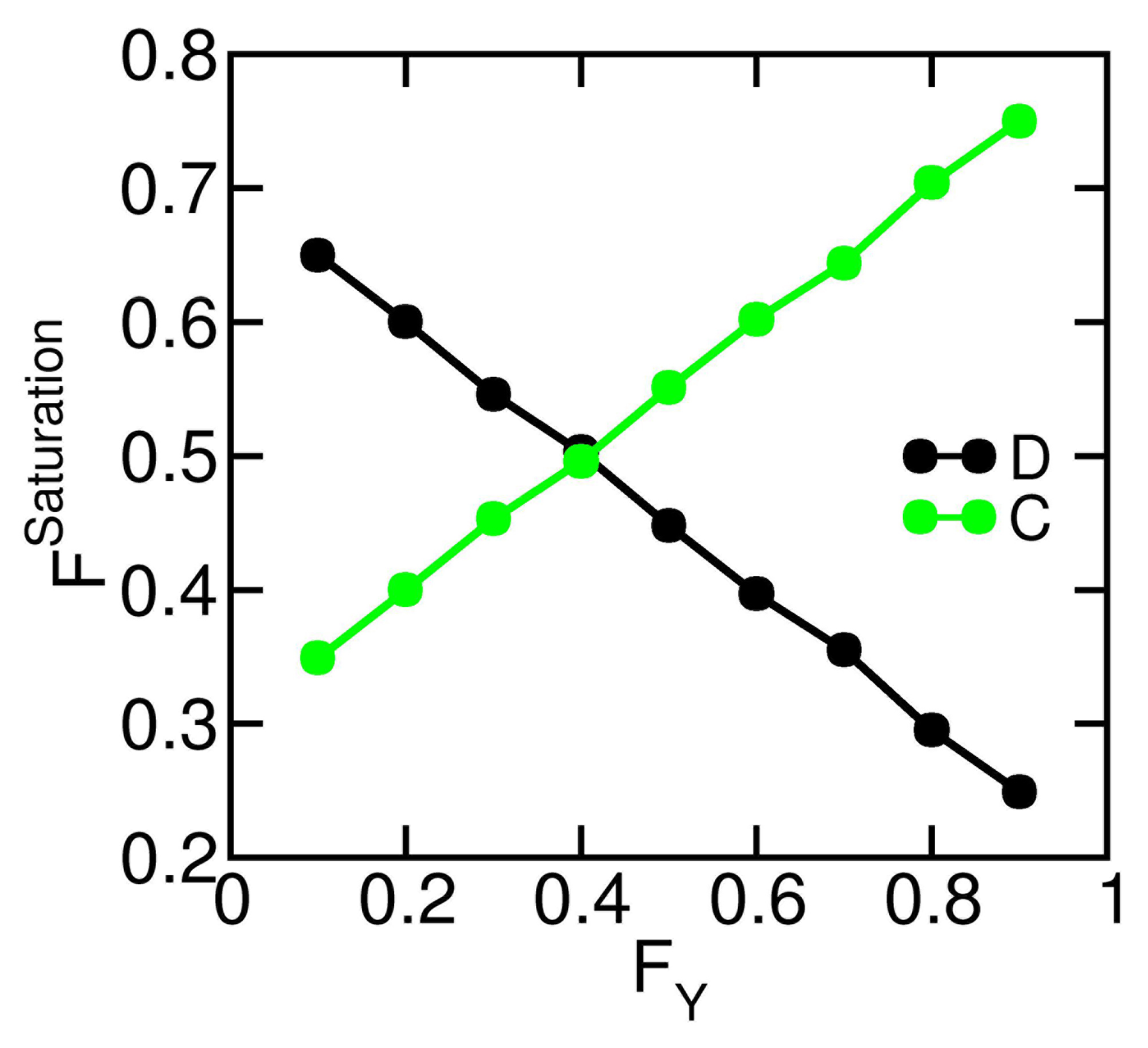}
 \caption{Dependence of the fraction of cured and dead population on the fraction of young population as obtained from CA simulations.}
 \label{fig8}
\end{figure}

In Fig. \ref{fig8} we plot the fraction of cured and dead population after temporal saturation from the plots shown in Fig. S3. We discover a linear variation for both with changing $F_Y$.   

\subsubsection{Duration of infection}

One of the most important parameters that determine the effectiveness of the spread of an epidemic is the time duration for which an individual remains infected before death or recovery. A person may or may not develop symptoms in this incubation period. For the latter case, it is most often not possible to determine whether a person has been infected or not. Hence, the probability of transmission from this section of the society is high.

Fixing the values of $P_R^Y$, $P_R^O$ and $A$ (Table \ref{tab1}), we vary the value of $P_I$ from 0.90 to 1.00 to study the time evolution of the different categories of the population (Fig. S4). This gives us an idea about the effect of the infection period for a particular disease. Here we neglect the effects of quarantine and social distancing to minimize the artefacts introduced by multiple parameters in the simulation. The probability of transmission $P_{Tr}$ is taken to be time-independent in this case ($P_{Tr}$ = 1.0).

Each of the above results are obtained by averaging over 100 individual simulations. The initial population density in each case is given by 2 \% of the total available land covered by susceptibles and 0.1 \% covered by infectives. The threshold age for defining young and old is taken to be 50, wherein 60 \% of the total population is considered to be young. $P_R^Y$ and $P_R^O$ are taken to be 0.8 and 0.3 respectively, considering that the young people have a greater chance of recovery. 

Fig. S4 a, b, c and d show the time evolution of S, I, C and D respectively for the different values of $P_I$. As expected, the rate of infection increases with an increase in the value of $P_I$. FS is found to decrease initially as $F_I$ increases. After a certain amount of time, $F_S$ attains saturation and $F_I$ reaches a maximum after which it decreases to 0 gradually. $F_C$ and $F_D$ finally take over after an initial increase and ultimately reach saturation.

\begin{figure}[H]
 \centering
 \includegraphics[width=3in,keepaspectratio=true]{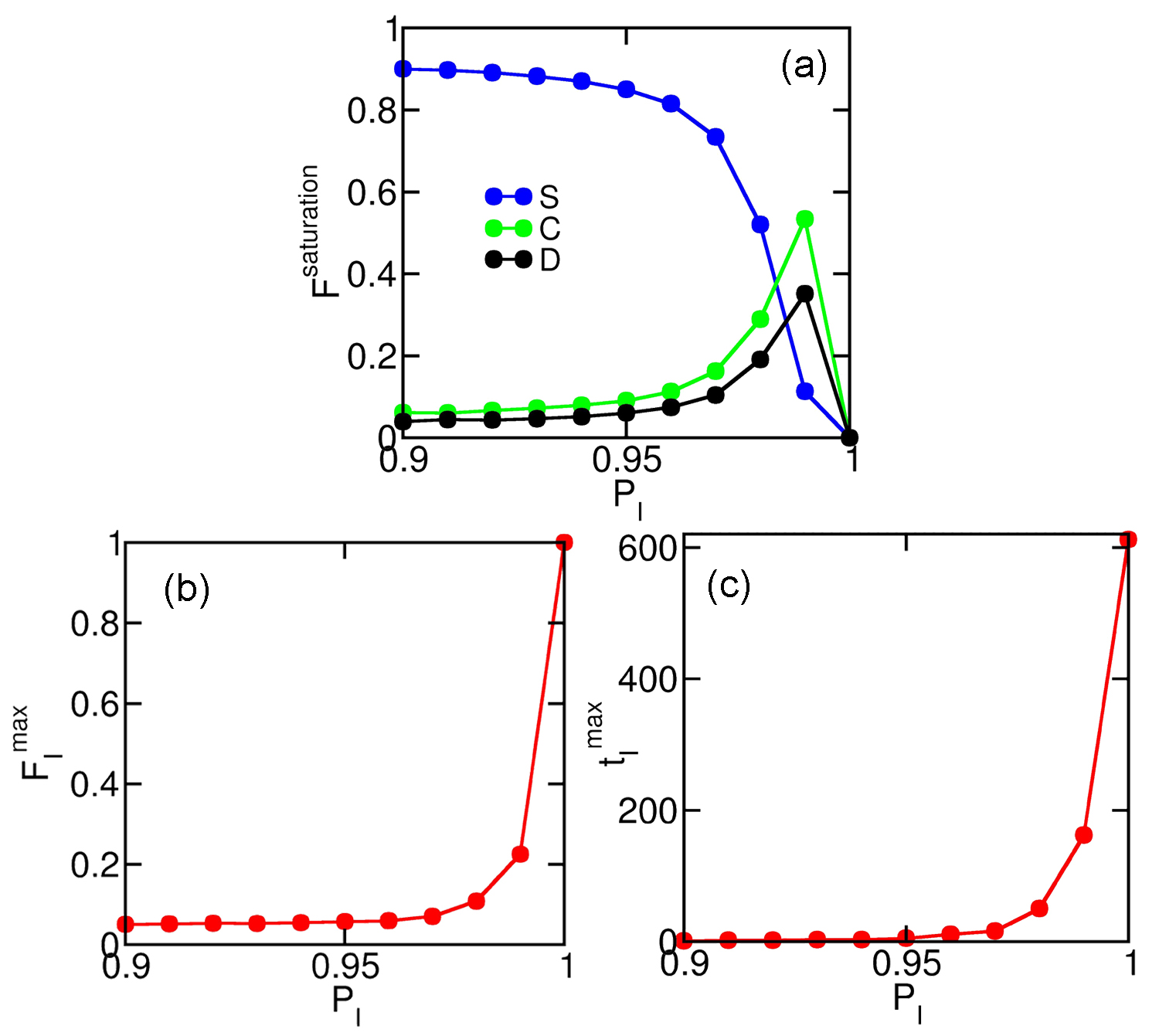}
 \caption{Dependence of the saturated values of (a) $F_S$, $F_C$, $F_D$, (b) peak height and (c) peak position of $F_I$ on $P_I$ obtained from CA simulations. A distinct non-linearity is observed at higher values of the control parameter.}
 \label{fig9}
\end{figure}

For $P_I$ = 1.0, the disease is continuously carried forward without any death or recovery. Hence, we do not have any data for $F_C$ and $F_D$ in this case. This however is unphysical. We plot the saturation values of $F_S$, $F_C$ and $F_D$ and the maxima height and position of $F_I$ with respect to the control parameter (which is $P_I$ in this case) in Fig. \ref{fig9}.

After an initial linear progression a non-linearity is observed in the fractions as the value of the control parameter is increased. We observe that the epidemic is more aggressive for $P_I > 0.97$. This is where the non-linearity sets in significantly. The exponential growth of $F_I$ and subsequent decrease in $F_S$ for higher values of $P_I$ denotes that for diseases in which death/recovery is seen after a long period of remaining infected are more likely to spread to a wider section of the community although the fatality rate might be low. This directly relates to the incubation period of the virus causing the disease. It should be noted that fatality rate and incubation period are two competing factors that determine the range of spread of a pandemic. Hence, for a virus to have a wide-spread effect like that of COVID-19, $P_I > 0.97$, while $P_D$ for the major fraction of the society is much lower.

\subsubsection{Quarantine and Social Distancing}

As a way to counter the spread of a pandemic, such as the present COVID-19 crisis, governments around the world have implemented nationwide lock down to some extent. Social distancing is being prescribed as an important tool that is expected to break the chain of virus spread and flatten the curve of infection. From our cellular automata simulations we quantify the importance of these two factors. Starting with the same population density, age distribution and recovery probability as mentioned in the previous section, a fraction ($F_Q$) of the susceptible part of the population is fixed in their positions throughout the simulation. This simulates the people who are quarantined or locked down in their homes. It is assumed that they do not go outside and has zero contact with infectives. Hence, they never get infected. These are immobile and remain healthy throughout the simulation. We take the value of $P_I$ to be 0.99 considering that the virus can spread widely. Taking the transmission probability ($P_{Tr}$) to be 1 (time-independent), we vary the value of $F_Q$ from 0.0 to 1.0 to see the effect of quarantine on the state of the population. $F_Q = 0$ means that no one is maintaining home quarantine, while $F_Q = 1$ denotes complete lock down of the entire healthy population.

\begin{figure}[ht]
 \centering
 \includegraphics[width=3in,keepaspectratio=true]{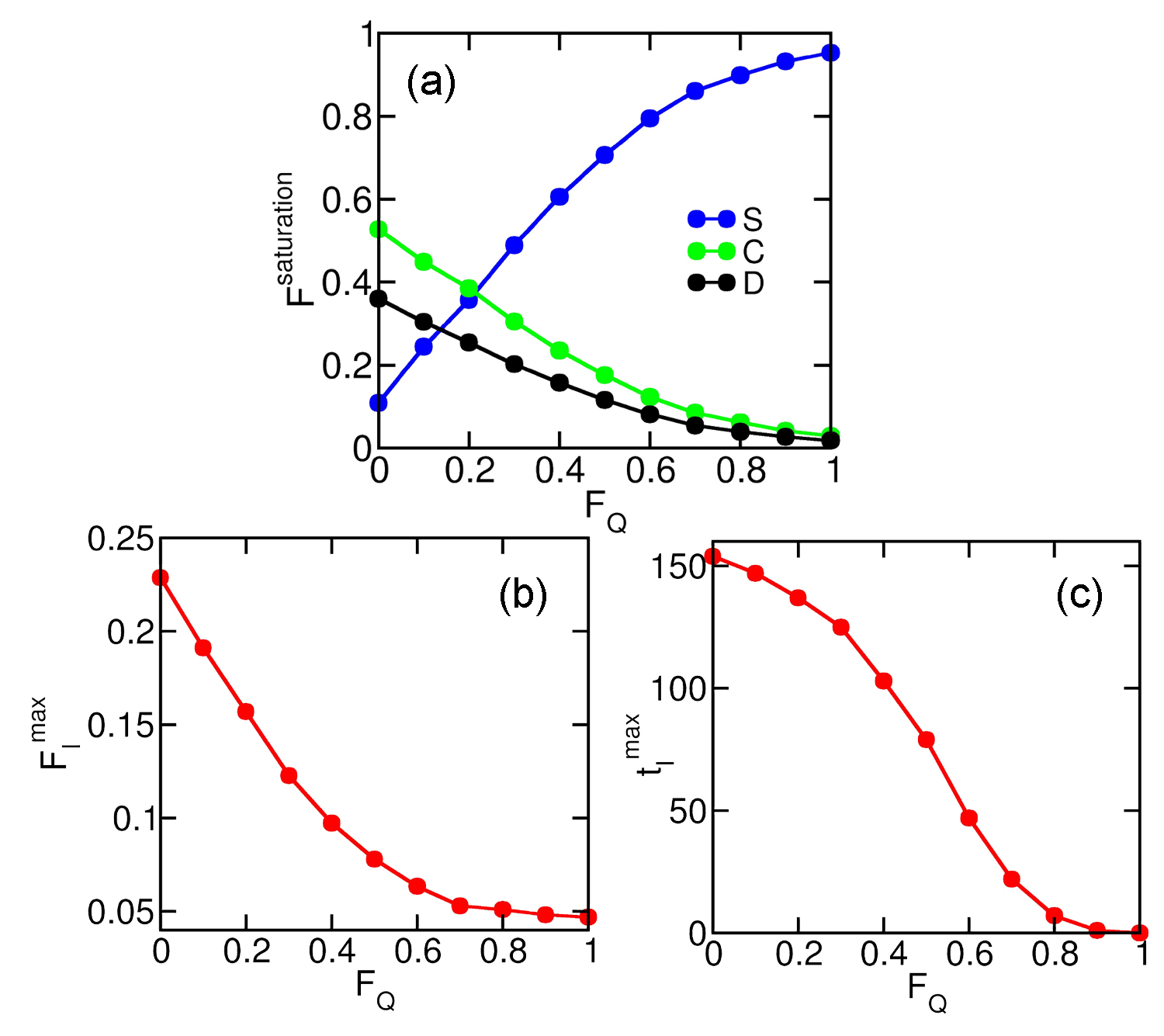}
 \caption{Dependence of the saturated values of (a) $F_S$, $F_C$, $F_D$, (b) peak height and (c) peak position of $F_I$ on $F_Q$ obtained from CA simulations. Higher quarantined fraction denotes lesser spread of infection.}
 \label{fig10}
\end{figure}

From our CA simulations we see that the fraction of people locking down themselves in their homes has a remarkable effect on the rate at which the virus spreads throughout the society (Fig. S5). The decay of the $F_S$ curve is much faster for lower values of $F_Q$. Also the infection maximum becomes flatter as the value of $F_Q$ is increased. The predicted rate of death is always lower than the rate of recovery. To understand these dependencies we plot the values of the fractions of S, C and D at which they become constant against the fraction of quarantined people in Fig. \ref{fig10}a. We also plot the height and position of the maxima of the fraction of infected people against the control parameter in Fig. \ref{fig10}b and c respectively. 

Examining Fig. \ref{fig10} the role of quarantine and lock down in defeating the spread of a pandemic becomes clear. For higher values of $F_Q$ the dependencies become non-linear. For $F_Q > 0.7$, the effect diminishes to a certain extent because of this non-linearity. Interestingly, the saturation values of $F_C$ and $F_D$ converge to almost 0 along with the onset of the said non-linearity. This denotes that with a decrease in the rate of infection the rates of both recovery and death become very low hinting towards a healthy overall society. The time taken by the infection to reach its peak is also significantly dependent of the factor of quarantine.

Though home quarantine is an effective measure to break the chain of a pandemic, it is not always feasible for everyone to strictly maintain the norms. For example, a person has to go to the market or medicine shop to obtain essential commodities for sustaining life at home. The doctors, nurses, police, sweepers and others who are actively involved in serving the society in the period of a pandemic also fall under this category. In addition, there are people who are indifferent to the seriousness of the situation and deliberately come out of their homes breaking the laws of lock-down put forward by the government. Hence a 100 \% quarantine ($F_Q = 1.0$) is unphysical. 

\begin{figure}[h]
 \centering
 \includegraphics[width=3in,keepaspectratio=true]{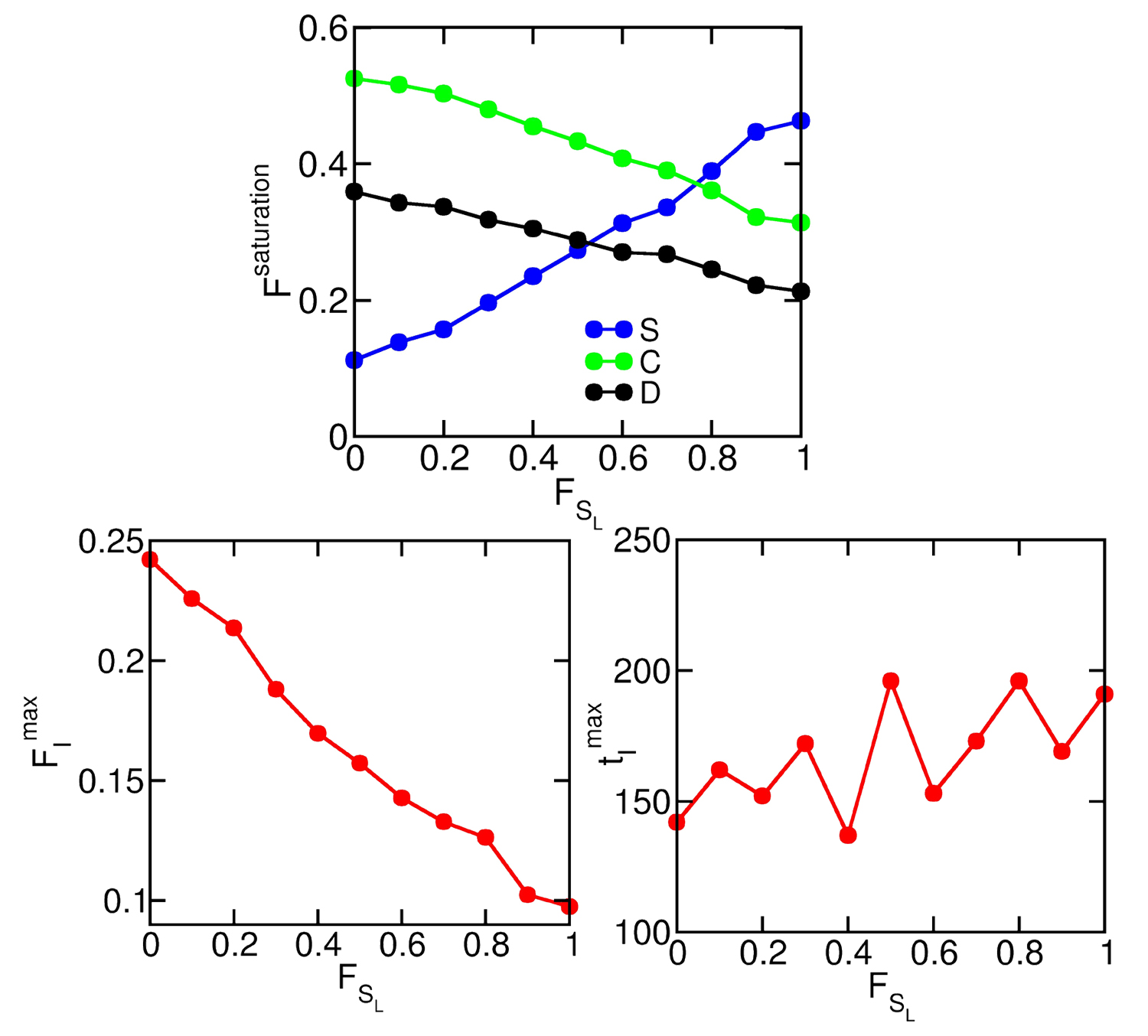}
 \caption{Dependence of the saturation values of (a) $F_S$, $F_C$, $F_D$, (b) peak height and (c) peak position of $F_I$ on $F_{SD}$ obtained from CA simulations.}
 \label{fig11}
\end{figure}

The people belonging to this ($1-F_Q$) fraction of the society try to avoid infection by maintaining social distancing and remaining careful about their surroundings. Hence, probability of transmission is drastically lowered for them. We vary the fraction of these individuals ($F_{SD}$) and monitor the change in the population with the spread of the disease. We consider the probability of transmission for the low susceptible people as $P_{Tr}^{SD}=0.2$ (much lesser than the general transmission probability, $P_{Tr} = 1.0$). As in the previous cases, the fractions of S, C and D attain a saturation after some time, while I reaches a maximum and decays thereafter (Fig. S6). We plot the respective saturation values, peak height and position as functions of $F_{SD}$ in Fig. \ref{fig11}.

In Fig. \ref{fig11}, the dependence of the time evolution of the different sections of the population is not as remarkable as in the previous case of quarantine (Fig. \ref{fig10}). Additionally, the time taken by the infection to reach the maximum is also weakly dependent on this control parameter. This weak dependence is observed even with a value of $P_{Tr}^{SD}$ as low as 0.2. For higher values of this probability, the dependencies are further lower. 

Hence, from our analysis, it becomes clear that to flatten the curve of infection, home quarantine is the most effective tool. Even with utmost caution while going out, the rate of infection can be high enough to affect a considerable section of the society in a smaller timescale.  

\section{Summary and Conclusion}

While COVID-19 has certain similarities with earlier epidemics (measles, small pox), there are certain remarkable features that distinguish the former from other pandemics. One important difference is that the immunity or the probability of getting infected depends on several factors that can be identified. Clearly, the accumulation of data across countries has become more accessible to general public and scientists alike. Secondly, the presence of a large number of asymptotic carriers which pose a serious threat to other less fortunate. In addition, there are certain similarities across the countries, as there are differences. Many of the emerging patterns are yet to be understood. It appears that modelling needs to be more generalized than those usually employed. There is an urgent need for understanding the dependencies.

The most common model used in modelling epidemics is the SIR model, already discussed here. We have discussed why this model might not be appropriate for the COVID-19 pandemic.
In this work we have employed two generalized schemes to model COVID-19 pandemic: (i)  a modified SIR model and (ii) a stochastic cellular automata simulations to understand the spatio-temporal evolution of a population undergoing a pandemic crisis. The key outcomes from the present study are as follows.

1.	The time evolution of the fractions of S, I and R in a population is dependent on the time-dependence of rate of infection. This is represented here in terms of $k_{S\rightarrow I}(t)$.  This in turn get translated to Similar dependence is also seen in the CA simulations where the transmission probabilities ($P_{Tr}$) are made time dependent. When a pandemic starts in a region (country for example) the nature of the spread is often unknown. Hence a proper protocol to control the spread and make the citizens cautious takes time. For this reason, the rate of transmission is higher in the beginning. With progress of time, this rate decreases as people learn more about the safety protocols. Sometimes, the rate may show oscillatory behaviour because of the heterogeneity of the population. Hence, time-dependence of the rate of transmission is an important factor.

2. 	From CA simulations, we find that both home quarantine and social distancing (for people coming out of their homes) are effective measures to fight the crisis of the spread of an infectious disease. However, even with utmost caution while out of home (low probability of getting infected), the peak of the infection curve do not show significant shift. On the other hand, the fraction of people quarantined strongly affects the position of the peak. Hence, staying at one’s home is the best policy during the spread of an infectious disease.

3.    Because of the conservation of the sum over sub-populations, the coupled equations become non-linear. This implies that an analysis of response to fluctuations in the control parameters might reveal interesting information. Such work is under progress.

4.   We also find that the incubation period of the virus plays an important role in determining the fate of the time progression of the pandemic. Longer the incubation period, (high probability of remaining infected) lesser are the chances that the infected individual will be symptomatic. Hence the disease in that person cannot be detected easily. As a result, this individual poses a severe threat of infection for the nearby susceptible population.

Other factors, like age distribution, initial population density of susceptibles, and time of lock-down of national borders are also major contributing issues that determine the spread of the disease. The rate constants, which are the parameters in the SIR model, represent average rates involving several factors (as mentioned above) that affect the overall temporal evolution of the population. We include these factors as parameters in terms of probabilities in the CA simulations. This not only gives us a better understanding of the roles of these individual factors but also allow us a more microscopic view into the physical map of a region affected by the spread of an infectious disease. However, it should be noted that both the model and the simulation produce limited results because of the inherent inhomogeneous nature of any pandemic and the limited availability of data. The models can be applied to small local geographical regions. However, they may fail to predict the outcome in a global picture.

\begin{acknowledgments}
We thank the Department of Science and Technology (DST, India) for partial support of this work. B.B. thanks Sir J. C. Bose fellowship for partial support. S.Mu. thanks DST, India for providing INSPIRE fellowship. S.Mo. thanks UGC, India for providing research fellowship.
\end{acknowledgments}

\bibliography{ref}
\end{document}